# I/O-efficient iterative matrix inversion with photonic integrated circuits

Minjia Chen[1], Yizhi Wang[1], Chunhui Yao[1], Adrian Wonfor[1], Shuai Yang[1], Richard Penty[1], Qixiang Cheng[1*]
1. Centre for Photonic Systems, Electrical Engineering Division, Department of Engineering, University of Cambridge, Cambridge, CB3 0FA, UK
*Corresponding author: qc223@cam.ac.uk*

## Abstract

Photonic integrated circuits have been extensively explored for optical processing with the aim of breaking the speed bottleneck of digital electronics. However, the input/output (IO) bottleneck remains one of the key barriers. Here we report a novel photonic iterative processor (PIP) for matrix-inversion-intensive applications. The direct reuse of inputted data in the optical domain unlocks the potential to break the IO bottleneck. We demonstrate notable IO advantages with a lossless PIP for real-valued matrix inversion and integral-differential equation solving, as well as a coherent PIP with optical loops integrated on-chip, enabling complex-valued computation and a net inversion time of 1.2 ns. Furthermore, we estimate at least an order of magnitude enhancement in IO efficiency of a PIP over photonic single-pass processors and the state-of-the-art electronic processors for reservoir training tasks and MIMO precoding tasks, indicating the huge potential of PIP technology in practical applications.

## Introduction

The input/output (IO) data movement in a processor typically has limited bandwidth and consumes considerable energy, leading to performance bottlenecks in processing systems that receive input data and store output data through an attached host with memory[1–4]. The IO issue becomes particularly severe when the problem size exceeds the capacity of the processor, a situation that is often encountered in practice. Thus, enhancing the computation-to-IO (C-to-IO) ratio becomes a key focus, i.e. increasing the number of implemented operations for a given input/output data size. In computation intensive tasks such as matrix multiplications and inversions, where multiple operations are performed on each piece of input data, the C-to-IO ratio can be improved by reusing the input data. The tensor processing unit (TPU) exemplifies this approach in accelerating neural network inference tasks. The TPU utilizes a systolic architecture to compute matrix multiplications in parallel by reusing the input data, demonstrating significant speed and energy efficiency improvements over central processing units (CPUs) and graphics processing units (GPUs)[2]. Nevertheless, the challenge of keeping pace with the ever-growing demands of compute-intensive applications has urged people to keep seeking faster and more energy-efficient solutions.

Photonic processors have long been considered as a promising alternative in algebra computations, owing to its inherent high parallelism and low energy consumption[4-7]. Photonic processors have shone through in the field of matrix multiplications, notably been demonstrated in voice recognition[8], image classification[9,10], and optical communication[11]. However, few research has been done on using photonic processors to accelerate matrix inversion, a more computationally expensive operation that is fundamental to many scientific and engineering problems such as numerical computations[12–16], statistics[12,17], wireless communication systems[18–22] and neural network training[23,24]. In most practical applications, the matrices to be inverted are diagonally dominant or sparse, and approximate inversion results are often sufficient for subsequent processing stages. Iterative inversion methods generally outperform direct methods in these scenarios since one can easily balance speed and accuracy by terminating the iterative process after several iterations, while interrupting the inversion process of direct methods midway yields incorrect results[12]. Classical iterative matrix inversion algorithms essentially perform matrix multiplications and additions iteratively. Implementing such algorithms on a widely reported photonic single-pass processor (PSP)[8,9,25,26] necessitates repetitive inputting and outputting of computation results in each iteration, thus generating heavy IO traffic.

In this paper, we report a novel photonic iterative processor (PIP) based on reconfigurable photonic integrated circuits for speeding up matrix inversion tasks. The inclusion of optical loopback enables iterative computations for direct matrix inverting with an enhanced I/O efficiency. The processor core is a matrix-vector multiplier comprising MZI units, which reduces the computation overhead by directly encoding the matrix elements on MZI arrays. Table 1 compares the C-to-IO ratios for inverting an $N \times N$ matrix using the iterative Richardson method[12] on four different platforms including the CPU, TPU, PSP, and our proposed PIP. The Richardson method is assumed to converge after $P + 1$ iterations. According to the table, the PIP exhibits the highest C-to-IO ratio of $N^3P/(2N^2 + N)$ among the four computing platforms, followed by the PIP, the TPU, and the CPU. In most application scenarios, $P$ is much larger than $N$, indicating an improvement of at least $N$ times in the C-to-IO ratio for PIPs compared to other platforms (See Supplementary 1 for more details).

**TABLE 1**

**A COMPARISON OF C-TO-IO RATIOS FOR INVERTING AN $N \times N$ MATRIX USING THE RICHARDSON METHOD**

| | **CPU** | **TPU** | **PSP** | **PIP** |
|---|---|---|---|---|
| Input data size | $2N^2 \cdot P$ | $N^2 \cdot P$ | $N^2 \cdot P$ | $N$ |
| Output data size | $N^2 \cdot P$ | $N^2 \cdot P$ | $N^2 \cdot P$ | $N^2$ |
| Memory access counts* | $3N^2 \cdot P$ | $N^2 \cdot (2P + 1)$ | $N^2 \cdot (2P + 1)$ | $2N^2 + N$ |
| Number of operations† | $N^3 \cdot P$ | $N^3 \cdot P$ | $N^3 \cdot P$ | $N^3 \cdot P$ |
| **C-to-IO ratio** | $N/3$ | $NP/(2P + 1)$ | $NP/(2P + 1)$ | $N^3P/(2N^2 + N)$ |

* This includes the number of times a processor needs to fetch input data, store output data, and load weight matrix (the matrix to be multiplied in each iteration). For CPU, the weight matrix is also fetched in a similar way as the input data and is therefore included in the input data size.
† Note the computations in the first iteration are neglected in the analyses for two reasons: (1) if the process converges for $P = 1$, then the matrix to be inverted is essentially an identity matrix whose inverse is itself and does not need to be computed. (2) In the first iteration, the algorithm essentially implements $N$ multiply-and-accumulate (MAC) operations, which can be neglected compared to the $N^2$ MAC operations in the following iterations.

We demonstrate, to the best of our knowledge, the first lossless reconfigurable PIP system that is capable of directly inverting matrices and solving integral and differential equations. Such a PIP system computes 4×4 real-valued matrix inversions with an accuracy of >97%, a C-to-IO ratio improvement of up to >7 times, a core energy efficiency of 4.6 MOPS/W (mega operations per second per watt), and a net inversion time of 2.6 µs, which is solely bounded by the length of fibre-based optical loops. The lossless PIP is then reconfigured to numerically solve real-valued integral and differential equations, reaching a mean absolute error of < 0.02, and up to >1.8 times C-to-IO ratio improvement. The first coherent PIP with on-chip optical loops is also demonstrated to break the loop-length limitation on the net inversion time. The coherent PIP is demonstrated to operate 2×2 complex-valued matrix inversions with an accuracy of > 98%, a core energy efficiency of 3.6 GOPS/W, and a net inversion time of 1.2 ns. Its enhancement in C-to-IO ratio reaches up to 2.8 times. Benefiting from the much-reduced IO demand, the proposed PIP is capable of reaching at least an order of magnitude IO efficiency enhancement compared with a single-pass optical processor and the state-of-the-art electronic processors, by emulating MIMO precoding tasks and reservoir training tasks. Our results indicate a promising way towards ever-powerful optical processors that could surpass IO limits.

## Results

### Photonic iterative processor architecture

The proposed PIP is tailored for matrix inversion problems that cannot be easily solved by traditional single-pass photonic processors. Solving integral and differential equations can be reduced to basic matrix computations including addition, subtraction, multiplication and inversion, which can all be solved optically by iterative algorithms using the PIP. Matrix inversion is computed through the Richardson method:

$$\boldsymbol{X}^{(k+1)} = (\boldsymbol{I}_N - \omega \boldsymbol{A})\boldsymbol{X}^{(k)} + \omega \boldsymbol{I}_N \ (k = 0,1,2, \dots) \quad (1)$$

where $\boldsymbol{A}$ is an $N \times N$ matrix operand to be inverted, $\boldsymbol{I}_N$ is the $N \times N$ identity matrix, $\omega$ is a parameter used to adjust the convergence of the inversion algorithm and the matrix operand is encoded in the weight bank via $\boldsymbol{I}_N$-$\omega\boldsymbol{A}$. $\boldsymbol{X}^{(k+1)}$ and $\boldsymbol{X}^{(k)}$ are output matrices after $k$+1 and $k$ iterations, and $\boldsymbol{X}^{(\mathbf{0})} = \omega\boldsymbol{I}_N$ is the initial input matrix that initiates the computation. The PIP generates matrix computation results one column at a time. Full-matrix inversion can be realised by using $N$ PIPs or by a multiplexing technique based on a different architecture we proposed[27].

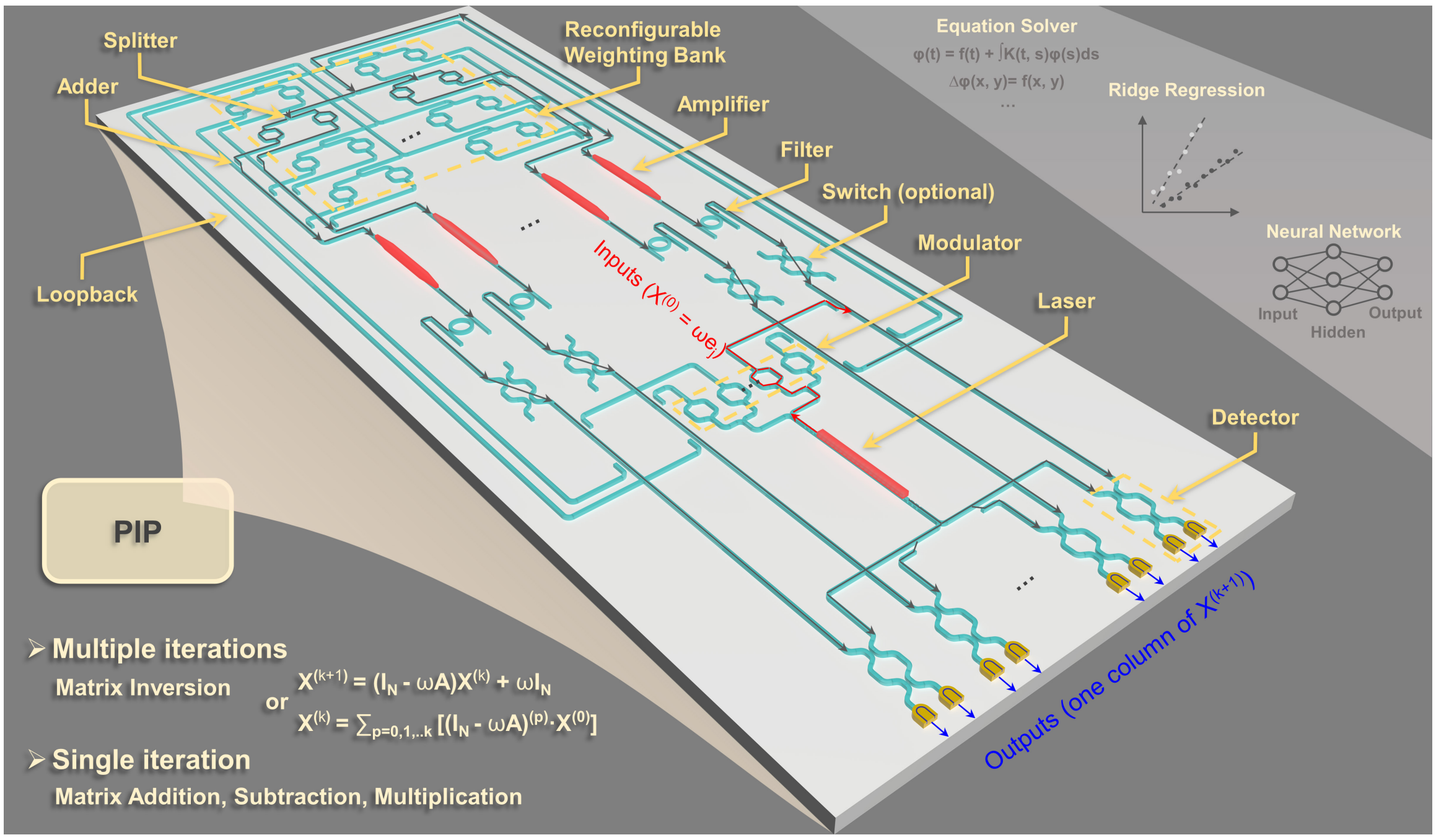

**Figure 1 | Conceptual figure of the photonic iterative processor (PIP).** Architecture of the proposed iterative photonic processor. The PIP serves as a photonic accelerator for inverting matrices which is widely used in equation solving, communication systems, robotics trajectory control, etc. Matrix inversion is solved by the Richardson method, whose results can be obtained by multiple iterations of the light signal in the PIP. Matrix addition, subtraction, and multiplication results can also be computed by a single iteration of the light signal in the PIP. For the lossless PIP system, the 4×4 reconfigurable weighting bank, adders and splitters are integrated on-chip. For the coherent PIP system, the 2×2 reconfigurable weighting bank, adders, splitters, optical switches, coherent detectors, and optical loopback are integrated on-chip. Monolithic integrations are possible and separately discussed in the section titled "Photonic integration techniques for a fully integrated PIP".

Fig. 1 depicts the architecture of the proposed PIP based on Richardson method, with grey arrows indicating signal flows. As highlighted by red arrows, the input pulse (representing the $j^{\text{th}}$ column of the initial input matrix, $\omega\boldsymbol{e}_j$) is generated by modulating the continuous wave (CW) laser in modulators. Upon entering the loop, part of the pulse is split and sent to detectors for outputs readout (one column of $\boldsymbol{X}^{(k+1)}$). The remaining part first passes an $N \times N$ weight bank with each MZI encoding one element of the $N \times N$ matrix, $\boldsymbol{I}_N$-$\omega\boldsymbol{A}$. Subsequently, the weighted pulses are summed by waveguide couplers to implement matrix-vector multiplication (MVM). The MVM results (one column of $(\boldsymbol{I}_N - \omega\boldsymbol{A})\boldsymbol{X}^{(k)}$) are then amplified to compensate for any loop loss. Optical filters are attached to remove excess amplified spontaneous emission (ASE) noise. The "clean" pulses now complete operations in one iteration and are either dropped or retained for recursive computing depending on the configuration of the optical switches. When the switches are "On", the "clean" pulses are sent for summation with the input pulse (between one column of $(\boldsymbol{I}_N - \omega\boldsymbol{A})\boldsymbol{X}^{(k)}$ and one column of $\omega\boldsymbol{I}_N$). Then part of the summed signals is split out and detected for outputs. The remaining parts enter the next circulation for iterative computation. Such recursive operation is terminated either when the optical switches are set to "Off" or when the input pulse ends. The PIP is also capable of computing matrix addition and multiplication in a single iteration by proper configuration. Detailed descriptions can be found in the supplementary 5.2.

## Real-valued matrix inversions

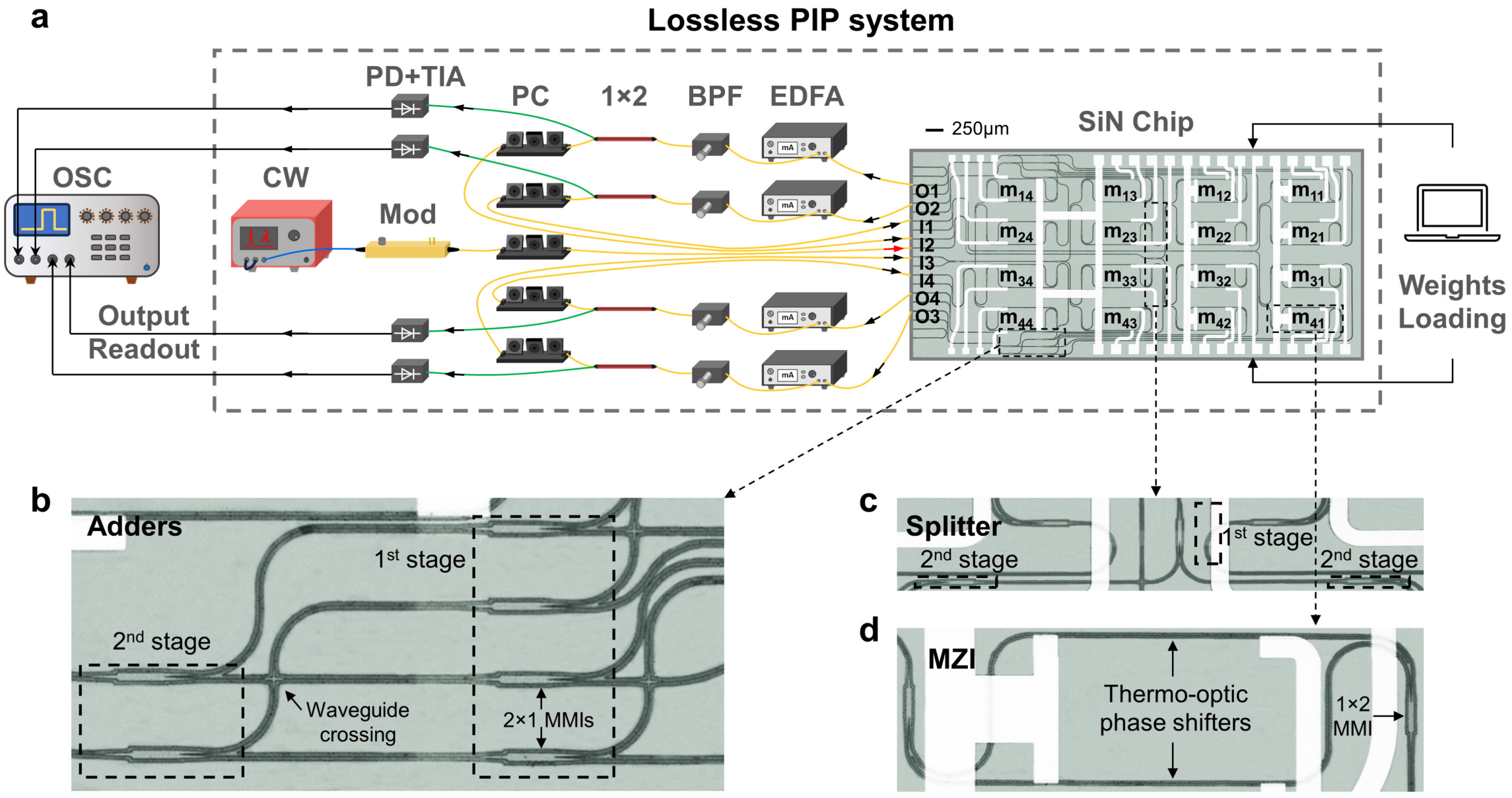

**Figure 2 |Lossless PIP system with a SiN chip core.** (a) Experimental set-up of the lossless PIP system. (b) Enlarged view of two adders comprising two stages of cascaded 2×1 multimode interferometers (MMIs). (c) Enlarged view of a splitter consisting of two stages of cascaded 1×2 MMIs. (d) Enlarged view of a weight unit comprising of a 1×1 thermo-optic (TO) MZI.

A 4×4 chip is taped out on the Silicon Nitride (SiN) platform, which is used as a processor core to form a lossless PIP system with off-the-shelf components as shown in Fig. 2a. The SiN chip integrates 4 adders, 4 splitters with a 4×4 sized MZI weighting bank, and their enlarged views are shown in Fig. 2b-d, respectively. Light is coupled in and out of the chip via edge couplers. Complete optical loopback paths are formulated by fibre components. The continuous-wave (CW) laser and optical modulator (Mod) correspond to the laser and modulator blocks in Fig. 1, generating an input vector $\omega \boldsymbol{e}_j$ that is coupled into the 4×4 chip. One column of the inverse results is computed each time by launching an optical input pulse to one input port and full matrix inversion is realized by sweeping different input ports (I1 – I4). The input pulse is first split into four copies on chip and then gets imprinted by the set of weights. The weighted signals are subsequently summed by 4 adders to perform an MVM, which are then coupled out of the chip and amplified by Erbium-doped fibre amplifiers (EDFAs), followed by bandpass filters (BPFs) to supress ASE noises. The 1×2 splitters allow part of the optical signals to be collected by the oscilloscope (OSC) after outputs readout in photodetectors (PDs) and transimpedance amplifiers (TIAs), while the remaining part are sent back to the chip via the optical loops. Polarisation controllers (PCs) are used to align the polarisation state of optical inputs to the chip. Electrical control is used for loading matrix weights, setting the modulator and operating outputs readout.

We use the set-up shown in Fig. 2a to demonstrate two real-valued matrix inversion examples (detailed matrix values, $\boldsymbol{A_1}$ and $\boldsymbol{A_2}$, can be found in Supplementary 4.1 and 4.2). As shown in Fig. 3a and Fig. 3d, $\boldsymbol{A_1}$ and $\boldsymbol{A_2}$ are loaded into the weighting bank as $\boldsymbol{M_1} = \boldsymbol{I_4} - \boldsymbol{A_1}$ and $\boldsymbol{M_2} = \boldsymbol{I_4} - \boldsymbol{A_2}$ respectively (See Methods and Supplementary for matrix weights characterisation). By injecting different unit vectors $\boldsymbol{X}_j^{(0)} = \boldsymbol{e}_j$ $(j = 1,2,3,4)$, different columns of the inversed matrices are obtained. Fig. 3b and Fig. 3e indicate a very good agreement between the ideal inverse results and the measured inverse results for the two inversion examples, respectively. The evolutions of inversion accuracy of $\boldsymbol{A_1}$ and $\boldsymbol{A_2}$ during convergence are traced and exhibited in Fig. 3c and Fig. 3f, reaching an inversion accuracy of 97.5% and 97.2% respectively.

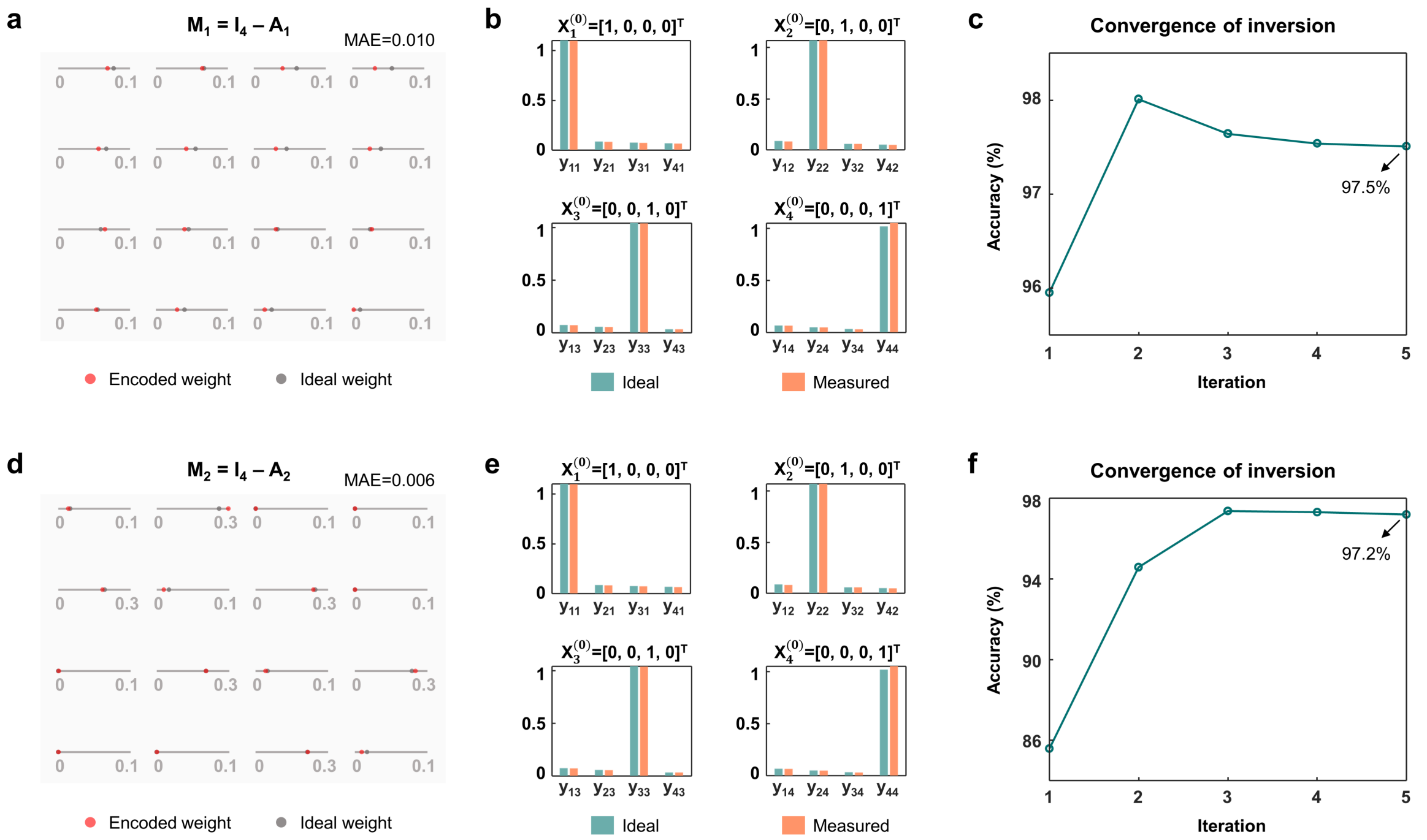

**Figure 3 | Real-valued matrix inversion examples.** (a) Ideal and encoded weight matrix $\boldsymbol{M_1}$. MAE: Mean absolute error. MAE = $\frac{1}{16}\sum_{j=1}^{4}\sum_{i=1}^{4}\left|\boldsymbol{M}^{encode}(i,j)-\boldsymbol{M}^{ideal}(i,j)\right|$. (b) Ideal and measured inverse matrix $\boldsymbol{A_1^{-1}}$. $\boldsymbol{X_j^{(0)}}$ is the $j^{th}$ column of the initial input matrix $\boldsymbol{X^{(0)}}$. (c) Evolution of inversion accuracy of $\boldsymbol{A_1}$ during convergence. Accuracy = $(1-\|\mathbf{A}_{\mathbf{meas}}^{-1}-\mathbf{A}_{\mathbf{ideal}}^{-1}\|/\|\mathbf{A}_{\mathbf{ideal}}^{-1}\|)\times 100\%$. (d) Ideal and encoded weight matrix $\boldsymbol{M_2}$. (e) Ideal and measured inverse matrix $\boldsymbol{A_2^{-1}}$. (f) Evolution of inversion accuracy of $\boldsymbol{A_2}$ during convergence.

## Solving real-valued integral and differential equation with matrix inversions

Integral and differential equations offer a powerful tool for quantifying the dynamics of systems that change over time or space, making them widely used in scientific research and engineering. Numerical solutions instead of analytical solutions are usually taken for real-world problems, which is essentially the problem of matrix inversion (See Method and Supplementary 3 and 5). Using the lossless PIP system that operates iteratively to solve integral and differential equations provides a novel computing paradigm that significantly reduces the demand of data movement. We adopt the system to solve an integral equation (IE, Fredholm integral equation of the second kind, Eq. (3a)), a second order ordinary differential equation (2$^{nd}$ order ODE, Eq. (3b)) with both using an 8-point discretization, and a partial differential equation (PDE, Poisson equation, Eq. (3c)) using a 4-point discretization.

$$f(t) = 1 + \int_{-1}^{1} 0.2\sqrt{t^2+s^2} f(s)ds, t \in [-1,1] \tag{3a}$$

$$d^2/dx^2\,[f(x)] - 2x \cdot d/dx\,[f(x)] - 50f(x) = -1, x \in [-1,1], f(-1) = f(1) = 1 \tag{3b}$$

$$\partial^2 u(x,y)/\partial x^2 + \partial^2 u(x,y)/\partial y^2 = -2\pi^2 \sin(\pi x)\sin(\pi y)\,, x,y \in [0,1], \partial_u = 0 \tag{3c}$$

An $N$-point discretization corresponds to $N \times N$ matrix inversions for IE and ODE, while it corresponds to $N^2 \times N^2$ matrix inversions for PDE. Block matrix computation techniques are employed to bridge the size of the problem and our chip, which could be readily mitigated by integrating a larger-scale processor on chip. Though using block matrix inversion techniques on a PIP increases the memory access count, the total memory access count and hence processing time is still much lower than that of traditional electronic processors and photonic single-pass processors (See Supplementary 6.4 for a detailed analysis). Figure 4a-c showcase the measured solutions based on the chosen discretization resolution, ideal solutions (from a conventional 64-bit digital computer) based on the chosen discretization resolution, and ideal solutions based on a finer discretization resolution, respectively. Mean absolute errors (MAEs $=\frac{1}{n}\sum_{j=1}^{n}\left|x_j^M - x_j^I\right|$, $\boldsymbol{x}^M$ is the measured solution, $\boldsymbol{x}^I$ is the ideal solution using the same discretization resolution) are shown at the top of each sub figure, indicating a very good agreement between the measured solutions and ideal solutions.

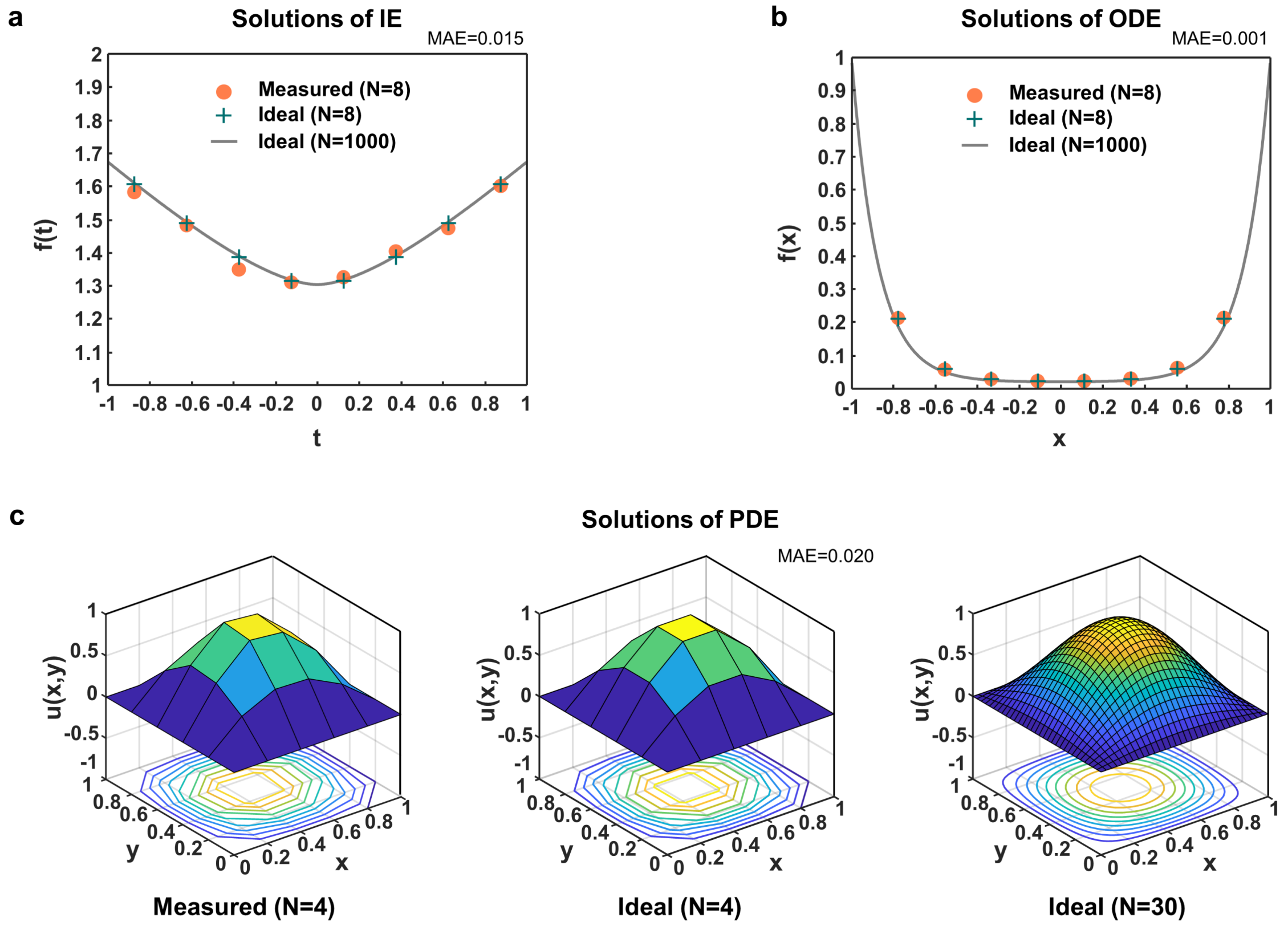

**Figure 4 | Solving real-valued integral and differential equations.** (a) Solutions to a Fredholm integral equation of the second kind. (b) Solutions to the 2nd order ordinary differential equation. (c) Solutions to the partial differential equation (Poisson equation). Mean absolute errors (MAEs) are indicated at the top of each sub figure.

## Complex-valued matrix inversions

One of the major applications of complex-valued matrix inversion is in MIMO decoding and precoding[19-21], which are fundamental to 5G/6G wireless communication systems. These algorithms require the inversion of complex-valued channel matrices, whose dimensions significantly increase owing to growing data demands from an increased number of end-users. Large scale complex-valued matrix inversion is computationally expensive and faces speed and power efficiency limitations in digital electronic computers. Unlike traditional methods that store a complex number as two separate real parts and process each part individually, photonic processors manipulate the amplitude and phase of optical signals simultaneously, enabling truly complex-valued computations with enhanced efficiency. Additionally, the PIP system can further improve the C-to-IO ratio and increase the inversion speed through its iterative optical loopback, significantly reducing IO access.

Here we show a coherent PIP system as shown in Fig. 5a for complex-valued matrix inversions. With optical loopback paths integrated on-chip, stable phase control can be achieved together with ultrafast processing time. Optical switches are integrated on-chip to facilitate device characterisation. Coherent outcomes are read out by off-chip balanced photodetectors (BPD) and captured in the OSC. EDFAs are used to compensate for coupling loss only and BPFs are used to remove excess ASE noise. We use the coherent PIP system to demonstrate three complex-valued matrix inversion examples (detailed matrix values, $\boldsymbol{A}_3$, $\boldsymbol{A}_4$ and $\boldsymbol{A}_5$, can be found in Supplementary 4.3 - 4.5). The measured and ideal inversion results for the three matrices including both the amplitude and the phase changes of each element of the inverse matrices are shown in Fig. 5b, 5d, and 5f respectively, indicating a good agreement with the ideal process. Figure 5c, 5e, and 5g shows the evolution of inversion accuracy of three matrices during multiple iterations. The inversion accuracy reaches 98.8%, 98.3% and 98.9% for $\boldsymbol{A_3}$, $\boldsymbol{A_4}$, and $\boldsymbol{A_5}$, respectively.

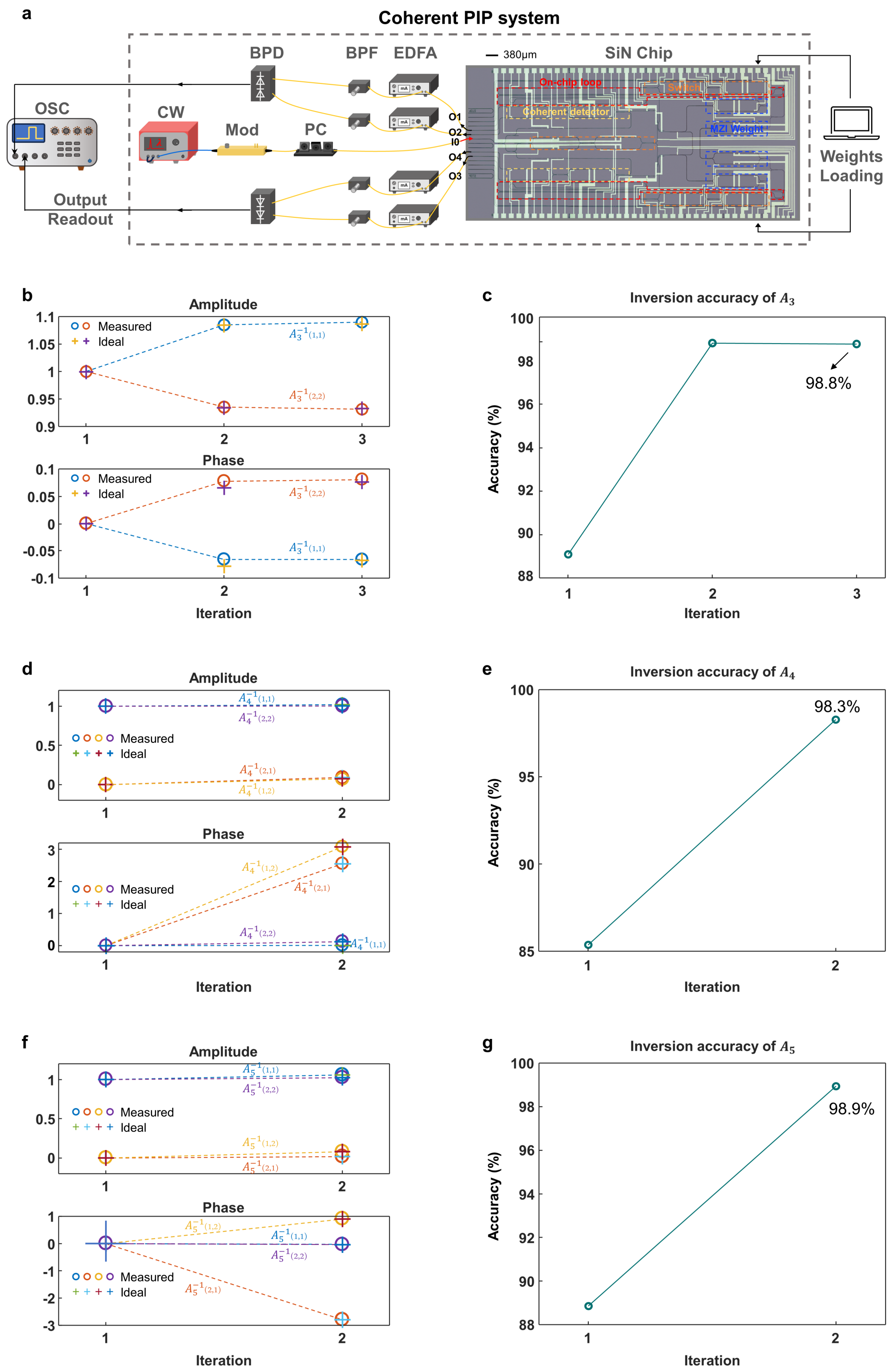

**Figure 5 | Demonstrated coherent PIP system and complex-valued matrix inversions.** (a) Experimental set-up of the coherent PIP system. (b) Ideal and measured inversion process of two diagonal elements of $\boldsymbol{A_3}$. (c) Evolution of inversion accuracy of $\boldsymbol{A_3}$ during convergence. (d) Ideal and measured inversion process of $\boldsymbol{A_4}$. (e) Evolution of inversion accuracy of $\boldsymbol{A_4}$ during convergence. (f) Ideal and measured inversion process of $\boldsymbol{A_5}$. (g) Evolution of inversion accuracy of $\boldsymbol{A_5}$ during convergence.

## IO advantages of the PIPs

We use saved processing time, $T_{save}$, saved energy consumption, $E_{save}$, improvement in total processing time, $\frac{t_{total_PSP}}{t_{total_PIP}}$, and improvement in C-to-IO ratio, $\frac{C_to_IO_{PIP}}{C_to_IO_{PSP}}$ to quantify the demonstrated IO advantages of the lossless PIP and the coherent PIP. The results are shown in Table 2. $q$ is the number of times a matrix is decomposed before computing on a PIP, $N$ is the PIP size, $t_{loop}$ is the single-iteration's processing time, and $P$ is the number of iterations for convergence. $T_0$ and $E_0$ are the processing time and energy consumption of a single IO access. Details about calculating each specs can be found in Supplementary 6.

**TABLE 2**
**ACHIEVED IO ADVANTAGES IN THE EXPERIMENTS**

| | Parameters | $T_{save}$ (μs) | $E_{save}$ (nJ) | $\frac{t_{total_PSP}}{t_{total_PIP}}$ | $\frac{C_to_IO_{PIP}}{C_to_IO_{PSP}}$ |
|---|---|---|---|---|---|
| $\boldsymbol{A}_1^{-1}$ | $q = 0, N = 4, t_{loop} = 130$ ns, $P = 5$ | 93.3 | 2.2 | 4.5 | 4.9 |
| $\boldsymbol{A}_2^{-1}$ | $q = 0, N = 4, t_{loop} = 130$ ns, $P = 8$ | 157.3 | 3.8 | 6.6 | 7.6 |
| IE | $q = 1, N = 4, t_{loop} = 130$ ns, $P_1 = 6, P_2 = 6$ | 250.7 | 6.0 | 1.7 | 1.75 |
| ODE | $q = 1, N = 4, t_{loop} = 130$ ns, $P_1 = 5, P_2 = 8$ | 272.0 | 6.5 | 1.8 | 1.82 |
| PDE | $q = 2, N = 4, t_{loop} = 130$ ns, $P_1 = P_3 = 9, P_2 = P_4 = 8$ | 736.0 | 17.7 | 1.2 | 1.2 |
| $\boldsymbol{A}_3^{-1}$ | $q = 0, N = 2, t_{loop} = 300$ ps, $P = 3$ | 12.0 | 0.29 | 2.8 | 2.8 |
| $\boldsymbol{A}_4^{-1}$ | $q = 0, N = 2, t_{loop} = 300$ ps, $P = 2$ | 6.7 | 0.16 | 2.0 | 2.0 |
| $\boldsymbol{A}_5^{-1}$ | $q = 0, N = 2, t_{loop} = 300$ ps, $P = 2$ | 6.7 | 0.16 | 2.0 | 2.0 |

According to the table, >7 times and >2 times improvement in the C-to-IO ratio are achieved for 4×4 real-valued matrix inversions and 2×2 complex-valued matrix inversions respectively. For solving the integral equation and the ordinary differential equation, 8×8 matrix inversions are performed on the 4×4 lossless PIP through one decomposition, still resulting in ~1.8 times improvement in IO efficiency. For solving the partial differential equation, 16×16 matrix inversions are performed on the 4×4 lossless PIP through two decompositions yet resulting in ~1.2 times improvement in IO efficiency. These results well verify the IO advantages of the PIP for matrix inversions.

# Discussion

## Processing time (latency) of an $N \times N$ PIP

The processing time of a PIP includes both the core processing time and the IO access time. The core processing time refers to the duration starting from when a signal is launched into the processor until the computation results are ready for acquisition, which is determined by the single-iteration's processing time, $t_{loop}$, and the number of iterations to reach convergence, $P$. The PIP system shown in Fig.1 generates the matrix inversion results one column a time, corresponding to a core processing time of $t_{core} = t_{loop} \cdot P \cdot N$. Using wavelength multiplexing techniques can reduce the core processing time to $t_{core_wdm} = t_{loop} \cdot P$. According to Table 2, the best demonstrated net inversion time, $t_{core}$, are 2.6 μs and 1.2 ns for the 4×4 and 2×2 cases, respectively. The loop length of the lossless PIP is mainly limited by the long fibre length (~35m) inside the EDFA, which can be removed by using semiconductor optical amplifiers (SOA) or on-chip gain components. It is worth noting that for the coherent PIP device, additional delay lines (~3.9 cm) are integrated as part of the loopback paths to ease the electrical readout, which can be shortened for a faster computation speed. To explore the future scaling of the PIP, we estimate the loop length and the single iteration's processing time of a fully integrated PIP with sizes ranging from 2×2 to 256×256 as shown in Fig. 6a-b. Both the loop length and the single iteration's processing time scale approximately linearly with PIP size.

The IO access time is defined as $t_{IO} = \frac{M \cdot B}{DR}$, where $M$ is memory access counts shown in Table 1, $B$ is the bit resolution of the data, and $DR$ is the data transfer rate. $t_{IO}$ includes the weight matrix loading time, input data fetching time, and output data storing time, and contributes to a heavy burden for digital electronic processors and single-pass photonic processors, while it contributes much less to the total inversion time for the PIP. The total processing (inversion) time of a PIP and a PSP can be estimated according to $t_{total_PIP} = t_{core_PIP} + t_{IO_PIP}$, and $t_{total_PSP} = t_{core_PSP} + t_{IO_PSP}$, respectively. More details can be found in Supplementary 7.2-7.3.

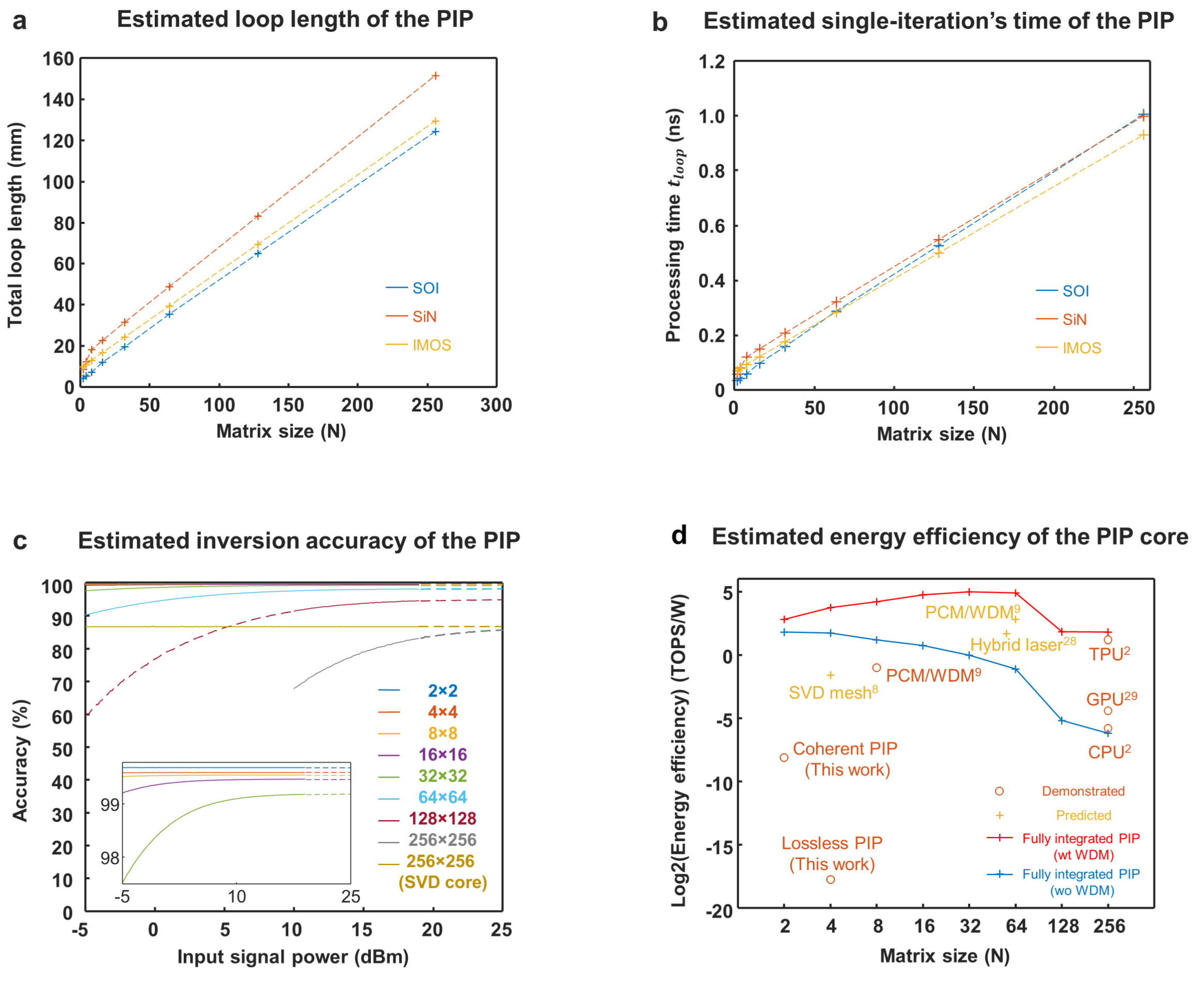

**Figure 6 | Performance analyses of the PIP.** (a) Estimated loop length of the PIP with sizes ranging from 2×2 to 256×256. Length of each component is based on the reported best values. (b) Estimated single-iteration's processing time with sizes ranging from 2×2 to 256×256. (c) Estimated inversion accuracies of the PIP with sizes ranging from 2×2 to 256×256. Note the 256×256 PIP with an SVD core only consider the 6 dB splitting and adding loss. (d) Estimated energy efficiency of the PIP core with sizes ranging from 2×2 to 256×256.

## Inversion accuracy of an $N \times N$ PIP

Inversion accuracy is defined in terms of matrix norm as: $Accuracy = (1 - \varepsilon) \times 100\% = (1 - \|\boldsymbol{A}_{meas}^{-1} - \boldsymbol{A}_{ideal}^{-1}\| / \|\boldsymbol{A}_{ideal}^{-1}\|) \times 100\%$, where $\varepsilon$ is the inversion error, $\boldsymbol{A}_{meas}^{-1}$ is the measured or simulated

inversion results, and $\boldsymbol{A}_{ideal}^{-1}$ is the ideal or theoretical inversion results calculated on a 64-bit traditional digital electronic computer. Three main error sources when performing matrix inversions on a fully integrated PIP include: 1) quantization error, 2) ASE noise introduced during amplification, and 3) thermal and shot noise introduced during detection. The estimated inversion accuracy of a PIP with sizes ranging from 2×2 to 256×256 is shown in Fig. 6c, indicating a relatively good inversion accuracy for a processor size of up to 256×256. More details can be found in Supplementary 7.4.

## Energy efficiency of an $\boldsymbol{N} \times \boldsymbol{N}$ PIP

Energy efficiency of the PIP core is defined as: $Energy\ efficiency = \frac{Number\ of\ operations/Second}{Energy\ consumption}$, which is the number of operations per second per energy consumption of the PIP core. The components consuming energy during the matrix inversion process include lasers, modulators, semiconductor optical amplifiers (SOAs), photodetectors (PDs), the MZI weight bank, and optical switches. The energy efficiency of the lossless PIP and the coherent PIP, together with the estimated energy efficiency of the PIP (with and without wavelength multiplexing techniques) and that of the state-of-the-art electronic and photonic processors are shown in Fig. 6d, indicating that the PIP structure is an energy-efficient architecture for matrix inversion tasks. More details can be found in Supplementary 7.5.

## Scalability of the PIP

In the above analyse, the PIP size is limited to 256×256. While this is sufficient for most real applications, with no more than 5 decompositions (see Supplementary 6for more details) ensuring the IO advantage for problem sizes up to 8192×8192, larger-scale PIPs are still preferable for tackling even more complex problems with more significant IO advantages.

The main constraints for further scaling up the PIP are similar to those for any large-scale photonic integrated circuit (PIC), i.e., the chip insertion loss and the available reticle size of the lithography. As shown in Fig. 6c, for PIPs larger than 128×128, good accuracy can still be achieved but requires high input signal powers, indicating that the round-loop loss is not well sustainable for higher-radix crossbar array-based MVM cores. A singular value decomposition (SVD)-based MVM core can be used to reduce signal splitting and addition losses, however, the accumulated insertion loss of the increased building blocks remains as a challenge. Figure 6c also reveals the estimated inversion accuracy of a 256×256 PIP with an SVD based MVM core, indicating good scalability. Nevertheless, the computation overhead for implementing SVD is non-negligible. Several approaches have been reported to effectively reduce the PIC insertion loss, including the use of a microelectromechanical-system (MEMS) MZI structure[30] and high-index doped silica glass waveguide[31]. Wafer-scale photonic integrated circuits are also demonstrated via inter-reticle waveguides[30], paving the way for very large-scale PIP systems.

## Envisioned IO advantages of the PIP for matrix-inversion-intensive applications

We further evaluate the IO advantages of our PIP in matrix-inversion-intensive problems, i.e., the MIMO precoding task and the reservoir training task which is essentially a ridge regression task. The MIMO precoding task solves $\hat{\mathbf{x}} = (\boldsymbol{H}^{\boldsymbol{H}}\boldsymbol{H} + \lambda\boldsymbol{I})^{-1}\boldsymbol{H}^{\boldsymbol{H}}\boldsymbol{y}$, while the reservoir training task solves $\boldsymbol{W}_{out} = \boldsymbol{Y}_d\mathbb{O}_{total}^T(\mathbb{O}_{total}\mathbb{O}_{total}^T + \lambda\boldsymbol{I})^{-1}$. The matrices to be inverted in these applications are typically diagonally dominant, and their feasibility of being inverted on a PIP is verified in the experiments. To predict the advantages of using a PIP for matrix inversions in these two applications, we offload the matrix inversion task to a hypothetical PIP based on the previous analyses of the PIP's processing time, accuracy, and energy consumption, and implement the main model on electronic processors. The estimated IO advantages of these two tasks are summarised in Table 3. More details about these two tasks can be found in Supplementary 8.

**TABLE 3**

**PREDICTED IO ADVANTAGES OF TWO PRACTICAL TASKS**

| | Parameters | $T_{save}$ | $E_{save}$ | $\frac{t_{total_PSP}}{t_{total_PIP}}$ |
|---|---|---|---|---|
| MIMO | $q = 0, N = 4, t_{loop} = 83.8$ ps, $P = 54, N_{samp} = 10^3 - 10^6$ *per day* | 0.15 ms-25 min per day | 27.3 μJ-273 J per day | 8.1 (43.9)* |
| MIMO | $q = 0, N = 8, t_{loop} = 124.7$ ps, $P = 104, N_{samp} = 10^3 - 10^6$ *per day* | 1.18 ms-3.3h per day | 211.8 μJ-2.1 kJ per day | 11.2 (90.9) |
| massive MIMO | $q = 0, N = 128, t_{loop} = 616.8$ ps, $P = 183, N_{samp} = 10^3 - 10^6$ *per day* | 0.53 s - 1472 h per day | 95.7 mJ-957 kJ per day | 31.6 (178) |
| massive MIMO | $q = 0, N = 256, t_{loop} = 1.1$ ns, $P = 186, N_{samp} = 10^3 - 10^6$ *per day* | 2.16 s - 6000 h per day | 0.39 J-3.9 MJ per day | 33.9 (181.6) |
| MNIST training | $q = 0, N = 10, t_{loop} = 94.3$ ps, $P = 14, N_{samp} = 6000, N_{epoch} = 10$ | 14.4 ms | 2.6 mJ | 8.5 (13.2) |

* Numbers in the brackets represent the cases when wavelength multiplexing techniques are used.

For a normal MIMO system or the current massive MIMO system, the size of the channel matrix is typically 4×4 and up to 8×8. For the future massive MIMO system, the channel matrix will be scaled to 128×128[32,33] or even larger. At least an order of magnitude IO improvement of the PIP is predicted for PIP sizes larger than 8×8. In practice, the MIMO channel matrices are time-varying and continuous matrix inversions are required. The updating period of a MIMO channel ranges from several minutes in a quasi-static environment to several microseconds in high-speed scenarios. This means ~$10^3$ to ~$10^{10}$ matrices need to be inverted per day, resulting in substantial amount of saved processing time and power consumption by using the PIP. For the reservoir training of the deep learning datasets, 8.5 times improvement in IO efficiency is predicted for a 10×10 PIP. Similarly, the total saved processing time and energy consumption becomes significant since a total of 60000 matrices needs to be inverted for the training task. The IO advantages of the PIP are more significant for datasets with more features. For both examples, wavelength multiplexing techniques can be used to enhance the IO advantages, as indicated by the numbers in the brackets in Table 3, leading to up to two orders of magnitude improvement in IO efficiency for MIMO precoding tasks and at least an order of magnitude improvement for the MNIST training task.

### Photonic integration techniques for a fully integrated PIP

According to Table 2, even if a PIP is not fully integrated on a single chip and with a limited scale, we can still gain notable benefits in speed and power consumption thanks to the reduced IO access counts. Moreover, as shown in Table 3, a fully integrated PIP as shown in Fig. 1 or a large-scale PIP makes the IO advantage more significant. Having optical gain in the loop supports a significantly increased number of iterations, making the PIP a more appealing technology in breaking the IO bottlenecks in practical matrix-inversion-intensive tasks. However, the integration of a light source or an amplifier on silicon chip still remains an active research area today, as silicon is an indirect bandgap material that precludes the possibility of a silicon laser. A number of hybrid and heterogeneous integration methods have been investigated to combine the power of III-V with the full capability of SOI. These include flip-chip bonding[34], die/wafer bonding[35,36], micro-transfer printing[37], direct epitaxial growth[38,39], optical wire bonding[40,41], and the membrane technology[42,43]. There are an increasing number of in-house demonstrations in the full integration of active and passive components, and when they are released to the public in the future, a fully integrated PIP can be realised immediately.

## Conclusion

In this paper, we propose a PIP with reconfigurable photonic circuits that is capable of handling signals in the optical domain recursively, achieving much higher computation-to-input/output ratio than traditional single-pass optical processors and digital electronic processors. This heralds a new photonic computing paradigm that significantly reduces the data shuttling cost. We showcase the first lossless PIP and the first coherent PIP with on-chip optical loops to demonstrate its power. High-fidelity optical computations including real-valued matrix inversions, real-valued integral and differential equation solving, and complex-valued matrix inversions are performed. By emulating the MIMO precoding tasks and the reservoir training tasks, the proposed PIP is shown to be capable of reaching at least an order of magnitude IO efficiency enhancement compared with a single-pass optical processor, benefiting from the much-reduced IO demand. Our work paves the way towards the next generation photonic processor that significantly enhances IO efficiency and processing speed.

## Methods

**Chip fabrication.** The photonic chip with a footprint of 2.8×6.6 $mm^2$ is fabricated on a SiN platform provided by CORNERSTONE multi-project wafer run using the standard deep ultraviolet lithography with a feature size of 250 nm. The platform comprises a 3 μm buried oxide layer, and a 2 μm silicon dioxide top cladding, and a 300 nm thick LPCVD SiN layer, which provides propagation loss of <1 dB/cm. Basic building blocks including the strip waveguide, the 1×2 MMI coupler, the 90° bend, the waveguide crossing and the edge coupler are customized using a commercial Lumerical FDTD simulator. The waveguide crossing has an extinction ratio of >30 dB. The edge coupler is based on a reverse taper structure, with a mode diameter of around 3.5 μm and a coupling loss of ~2.5 dB per facet.

**Chip characterisation (lossless PIP).** 16 MZI weights are characterised independently by launching a laser into the chip and measuring the output light intensity while sweeping the applied voltage to the heaters on the MZI arms. 16 transmission-voltage (T-V) or transmission-power (T-P) curves are recorded and fitted to form lookup tables for loading matrix weights. The effective weight of each matrix element is a combination of the attenuation in the weighting bank, the loss in the loop including the MZI unit, and the gain in that loop. This is achieved by creating an optical loop between input and output waveguides for a single MZI. The effective weight is then determined by launching an optical pulse into the loop and measuring the ratio of subsequent output pulses. Note that there is an EDFA within the loop whose purpose is to compensate for optical losses in the loop. Finally the attenuation in the MZI unit is adjusted to provide the required attenuation for each matrix element.

**Chip characterisation (coherent PIP).** In order to characterise the 4 MZI weights, two loop switches, one input switch, and two coherent detectors need to be characterised first. The "On/Off (Bar/Cross)" states of two loop switches are characterised by measuring the transmission of a light signal through two edge-coupled ports of the switch while sweeping the applied voltage to the heaters on the MZI arms. Configuring the loop switches to "On/Off (Bar/Cross)" state corresponds to maintaining/terminating the iterative process. The "Bar/Cross" state of the input switch, which corresponds to injecting unit vectors to two input ports, is characterised with both loop switches configured to "Off" states. Then the two coherent detectors, each comprising an MZI, are characterized respectively by setting the input switch to either "Bar" or "Cross" state and measuring the photo-detected electrical signals with both loop switches in "Off" states. The effective attenuation and phase shift of each MZI weight unit is the total attenuation and phase shift within the loop respectively, which are determined by launching optical pulses into two input ports respectively and measuring the ratio of the first two pulses.

**Experimental setup.** The light source (Thorlabs TLX1 tunable laser) is set to a 1550 nm wavelength and 5 dBm output power. Manual polarisation controllers are used to align light polarisations to the chip. The optical switch is Thorlabs LNA6213 intensity modulator. Amplifiers are EDFAs from Connect Laser which can provide >35 dB gain. BPFs are filters from WL Photonics with 0.1 nm 3 dB bandwidth. Outputs are recorded in a 4-channel Keysight DSO-S 404A oscilloscope. The system requires electrical control. A customized Matlab program is used to characterize the chip. The switch is controlled by a Tektronix AFG3102C function generator. Outputs from the oscilloscope are sent back to an electronic computer for analysis.

**Numerical methods.** Discretization is needed for numerically solving equations (see Supplementary 2). The rectangular integration technique is used for solving integral equations by approximating the integral by summing a series of rectangular partitions under the curve. Fredholm integral equations of the second kind can be written

as

$$f(t_i) = c(t_i) + \frac{b-a}{N} \cdot \sum_{j=1}^{N} K(t_i, s_j) f(s_j). \tag{6}$$

$[a, b] = [-1, 1]$ is the integral interval. $N = 8$ is the number of equally divided subintervals of $[a, b]$. Assume the divisions are the same along both $t$ and $s$ axes. $t_i$ and $s_j$ $(i, j = 1, 2, \dots, N)$ are midpoints of the subintervals along both axes (called the discretized points). $K(t_i, s_j)$ is the value of kernel function at the discretized point $(t_i, s_j)$. $f(t_i)$ is the value of the function to be solved at discretized points $t_i$. $c(t_i)$ is the value of the input function at discretized points $t_i$. Eq. (2) can be expressed in a matrix form as $\boldsymbol{f} = \boldsymbol{c} + \frac{b-a}{N}\boldsymbol{K}\boldsymbol{f}$. The solution of the linear equation is $\boldsymbol{f} = (\boldsymbol{I} - \frac{b-a}{N}\boldsymbol{K})^{-1}\boldsymbol{c}$.

The finite difference method is used to solve differential equations by using finite difference formulas at evenly spaced grid points to approximate the differential equations. There are three types of difference formulas, which are central, forward and backward differences. Here we use central difference to approximate the equations. The first and second order derivatives of ODEs (1D system) can be written as

$$\frac{dy}{dx} = \frac{y_{i+1} - y_{i-1}}{2h} \tag{7}$$

$$\frac{d^2y}{dx^2} = \frac{y_{i+1} - 2y_i + y_{i-1}}{h^2} \tag{8}$$

where $i$ is the index of the desired grid point, $i$−1 and $i$+1 are the indices of the neighbouring points, and $h$ is the grid size. A fixed grid size is used for simplicity. For PDEs (2D system), the second order partial derivatives with respect to variable $x$ and the gradient relationship can be written as

$$\frac{\partial^2 u(x,y)}{\partial x^2} = \frac{u_{i+1,j} - 2u_{i,j} + u_{i-1,j}}{h^2} \tag{9}$$

$$\Delta u(x, y) = \frac{\partial^2 u(x,y)}{\partial x^2} + \frac{\partial^2 u(x,y)}{\partial y^2} = \frac{u_{i+1,j} + u_{i,j+1} - 4u_{i,j} + u_{i-1,j} + u_{i,j-1}}{h^2} \tag{10}$$

The discretized equations (6)-(10) are then mapped into coefficient matrices which describes the relation between a point and other points in the grid. Solving differential equations are then converted to solving matrix inversions.

## Acknowledgement

This work was supported by the European Union's Horizon 2020 research and innovation programme, project INSPIRE, and UK EPSRC, project QUDOS (EP/T028475/1). The authors thank CORNERSTONE for providing free access to their second SiN MPW run (funded by the CORNERSTONE 2 project under Grant EP/T019697/1). The authors also thank Prof. Hui Zhang, Dr. Mark Holm, and Mr. Zexing Li for helpful discussions.

### Author contributions

M.C. and Q.C. conceived the idea. M.C. performed numerical simulations, designed the schematics of the photonic chips, built the experimental set-ups, conducted the experiments with helpful advice from A.W., S.Y. and Q.C., and analysed the results. Y.W. helped with the coherent PIP test and conducted simulations on using the PIP for ridge regression and MNIST training. C.Y. and M.C. designed the layouts of the photonic chips. M.C. and Q.C. wrote the manuscript, with inputs from all authors. Q.C. and R.P. supervised the project.

### Data availability

Additional data related to this publication is available at https://doi.org/10.17863/CAM.????..

### Code availability

The code used in this study is available from the corresponding author upon request.

### Competing interests

The authors declare no competing interests.

## References

1. Kung, H. T. Why systolic architectures? *MC* **15**, 37–46 (1982).
2. Jouppi, N. P. *et al.* In-Datacenter Performance Analysis of a Tensor Processing Unit. in *Proceedings of the 44th Annual International Symposium on Computer Architecture* 1–12 (Association for Computing Machinery, New York, NY, USA, 2017).
3. Alexoudi, T. *et al.* Optics in Computing: From Photonic Network-on-Chip to Chip-to-Chip Interconnects and Disintegrated Architectures. *JOURNAL OF LIGHTWAVE TECHNOLOGY* **37**, (2019).
4. McMahon, P. L. The physics of optical computing. *Nat Rev Phys* 1–18 (2023).
5. Ambs, P. Optical Computing: A 60-Year Adventure. *Advances in Optical Technologies* **2010**, 1–15 (2010).
6. Solli, D. R. & Jalali, B. Analog optical computing. *Nature Photon* **9**, 704–706 (2015).
7. Athale, R. & Psaltis, D. Optical Computing: Past and Future. *Optics & Photonics News, OPN* **27**, 32–39 (2016).
8. Shen, Y. *et al.* Deep learning with coherent nanophotonic circuits. *Nature Photon* **11**, 441–446 (2017).
9. Feldmann, J. *et al.* Parallel convolutional processing using an integrated photonic tensor core. *Nature* **589**, 52–58 (2021).
10. Ashtiani, F. *et al*. An on-chip photonic deep neural network for image classification. *Nature* **606**, 501–506 (2022).
11. SeyedinNavadeh, S. *et al.* Determining the optimal communication channels of arbitrary optical systems using integrated photonic processors. *Nat. Photon.* **18**, 149–155 (2024).
12. Watkins, D. S. *Fundamentals of Matrix Computations*. (Wiley-Interscience, New York, 2002).
13. Palmer, T. Modelling: Build imprecise supercomputers. *Nature* **526**, 32–33 (2015).
14. Altrock, P. M. *et al*. The mathematics of cancer: integrating quantitative models. *Nat Rev Cancer* **15**, 730–745 (2015).
15. Achdou, Y. *et al*. Partial differential equation models in macroeconomics. *Philosophical Transactions of the Royal Society A: Mathematical, Physical and Engineering Sciences* **372**, 20130397 (2014).
16. Bauer, P. *et al*. The quiet revolution of numerical weather prediction. *Nature* **525**, 47–55 (2015).
17. Montgomery, D. C. *et al*. *Introduction to Linear Regression Analysis*. (John Wiley & Sons, 2021).
18. Hashima, S. & Muta, O. Fast matrix inversion methods based on Chebyshev and Newton iterations for zero forcing precoding in massive MIMO systems. *J Wireless Com Network* **2020**, 34 (2020).
19. Rosário, F. *et al*. Fast Matrix Inversion Updates for Massive MIMO Detection and Precoding. *IEEE Signal Processing Letters* **23**, 75–79 (2016).
20. Prabhu, H. *et al.* Hardware efficient approximative matrix inversion for linear pre-coding in massive MIMO. in *2014 IEEE International Symposium on Circuits and Systems (ISCAS)* 1700–1703 (2014).
21. Tang, C. *et al*. High Precision Low Complexity Matrix Inversion Based on Newton Iteration for Data Detection in the Massive MIMO. *IEEE Communications Letters* **20**, 490–493 (2016).
22. Zhang, C. *et al.* On the Low-Complexity, Hardware-Friendly Tridiagonal Matrix Inversion for Correlated Massive MIMO Systems. *IEEE Transactions on Vehicular Technology* **68**, 6272–6285 (2019).
23. Tanaka, G. *et al.* Recent advances in physical reservoir computing: A review. *Neural Networks* **115**, 100–123 (2019).
24. Lukoševičius, M. & Jaeger, H. Reservoir computing approaches to recurrent neural network training. *Computer Science Review* **3**, 127–149 (2009).
25. Zhou, H. *et al.* Photonic matrix multiplication lights up photonic accelerator and beyond. *Light Sci Appl* **11**, 30 (2022).
26. Zhang, H. *et al.* An optical neural chip for implementing complex-valued neural network. *Nat Commun* **12**, 457 (2021).
27. Chen, M. *et al.* Iterative photonic processor for fast complex-valued matrix inversion. *Photon. Res.* **10**, 2488 (2022).
28. Nahmias, M. A. *et al.* Photonic Multiply-Accumulate Operations for Neural Networks. *IEEE JOURNAL OF SELECTED TOPICS IN QUANTUM ELECTRONICS* **26**, (2020).
29. NVIDIA Tesla P40 Specs. *TechPowerUp* https://www.techpowerup.com/gpu-specs/tesla-p40.c2878 (2024).
30. Seok, T. J. *et al.* Wafer-scale silicon photonic switches beyond die size limit. *Optica, OPTICA* **6**, 490–494 (2019).
31. Zhu, X. *et al.* Low-loss and polarization insensitive 32 × 4 optical switch for ROADM applications. *Light Sci Appl* **13**, 94 (2024).
32. Larsson, E. G. *et al.* Massive MIMO for next generation wireless systems. *IEEE Communications Magazine* **52**, 186–195 (2014).
33. Lu, L. *et al.* An Overview of Massive MIMO: Benefits and Challenges. *IEEE Journal of Selected Topics in Signal Processing* **8**, 742–758 (2014).
34. Lin, S. *et al.* Efficient, tunable flip-chip-integrated III-V/Si hybrid external-cavity laser array. *Opt. Express, OE* **24**, 21454–21462 (2016).
35. Roelkens, G. *et al.* III-V/Si photonics by die-to-wafer bonding. *Materials Today* **10**, 36–43 (2007).
36. Liang, D. *et al.* Low-Temperature, Strong SiO2-SiO2 Covalent Wafer Bonding for III–V Compound Semiconductors-to-Silicon Photonic Integrated Circuits. *J. Electron. Mater.,* **37**, 1552–1559 (2008).
37. Zhang, J. *et al.* III-V-on-Si photonic integrated circuits realized using micro-transfer-printing. *APL Photonics* **4**, 110803 (2019).
38. Shi, Y. *et al.* Optical pumped InGaAs/GaAs nano-ridge laser epitaxially grown on a standard 300-mm Si wafer. *Optica, OPTICA* **4**, 1468–1473 (2017).
39. Scherrer, M. *et al.* In-Plane Monolithic Integration of Scaled III-V Photonic Devices. *Applied Sciences* **11**, 1887 (2021).
40. Gu, Z. *et al.* Optical transmission between III-V chips on Si using photonic wire bonding. *Opt. Express, OE* **23**, 22394–22403 (2015).
41. Billah, M. R. *et al.* Hybrid integration of silicon photonics circuits and InP lasers by photonic wire bonding. *Optica, OPTICA* **5**, 876–883 (2018)
42. Matsuo, S. *et al.* Directly modulated buried heterostructure DFB laser on $SiO_2$/Si substrate fabricated by regrowth of InP using bonded active layer. *Opt. Express, OE* **22**, 12139–12147 (2014).
43. Jiao, Y. *et al.* Indium Phosphide Membrane Nanophotonic Integrated Circuits on Silicon. *Phys. Status Solidi A* **217**, 1900606 (2020).